\documentclass[aps,prb,twocolumn,showpacs,amsmath,amssymb,floats]{revtex4-1}

\usepackage{graphicx}
\usepackage{dcolumn}
\usepackage{bm}

\begin{document}

\title{Photocurrent and photoacoustic  detection of plasmonic behavior of CdSe quantum dots grown in Au nanogaps}

\author{Andriy Nadtochiy}%
\affiliation{
Faculty of Physics, Taras Shevchenko Kyiv National University, Kyiv 01601, Ukraine
}%

\author{Om Krishna Suwal}%
\affiliation{%
Department of Physics and Astronomy, Seoul National University, Seoul 08826, Republic of Korea
}%

\author{Dai-Sik Kim}%
\affiliation{%
Department of Physics, Quantum Photonics Institute and Center for Atom Scale Electromagnetism, Ulsan National Institute of Science and Technology, Ulsan 44919, Republic of Korea
}%

\author{Oleg Korotchenkov}%
\affiliation{
Faculty of Physics, Taras Shevchenko Kyiv National University, Kyiv 01601, Ukraine
}%
\affiliation{
Erwin Schr\"{o}dinger International Institute for Mathematics and Physics, University of Vienna, Vienna 1090, Austria
}%

\begin{abstract}
    In this work, the influence of Au plasmonics on photocurrent generation in the visible wavelength range in integrated thiol-linked CdSe quantum dot/Au nanogap structures is demonstrated. The plasmonic absorption is sensed utilizing photoacoustic  (PA) detection technique. The laser diode is modulated to generate the PA excitation that oscillates the air-filled cell. When light absorption increases, an enhanced acoustic signal is captured by a microphone. The observed enhancement in the PA response is related to plasmonic absorption by the Au layers and the response is further enhanced by about 20\% due to CdSe QDs. In our structure, the surface plasmon resonance (SPR) wavelength is approximately 500 nm. The SPR is utilized for generating photocurrents in CdSe quantum dots. Due to energy transfer from the dot to closely spaced Au surface through thiol links, a smooth transmission channel of electrons is established that forms a detectable photocurrent, which can be tuned by a bias voltage. These plasmonic nanogap structures can enable higher sensitivity in photovoltaics, photodetection and sensing.

\end{abstract}

\maketitle

\newpage

\section{Introduction}
Optical properties of metallic nanostructures are determined by the collective excitations of their conduction electrons with respect to the positive background ions, which are called plasmons.\cite{Kreibig} Plasmonic nanostructures are widely based on Ag and Au, which exhibit localized surface plasmon resonance as a collective dipolar oscillation of free electrons in phase with the incident electromagnetic field. Plasmonic nanostructures have a wide variety of applications. \cite{Caucheteur,Choi,Hou,Mandal} If plasmons on neighboring parts of the nanostructures joined by surfaces and interfaces interact, they can hybridize just like the electron wave functions of molecular orbitals. Such plasmon hybridization represents a powerful paradigm for designing metallic resonant nanostructures.\cite{Prodan,Laghrissi} The resulting optical properties of plasmonic nanostructures are in many ways complementary to those of semiconductor quantum dot nanostructures.\cite{Wang,Kim,Mishra,Smith} 

Metallic nanogaps have been advantageous in offering interesting quantum plasmonics opportunities.\cite{Esteban} It has been observed that a metal nanogap can highly confine an electric field of the incident long-wavelength light between the gap of two metal surfaces (referred to as the field enhancement).\cite{Im,Chen,Siegfried} A metal/insulator/metal nanogap offers effective mode volumes well below the diffraction limit in the gap material, even if there is a significant loss of energy inside the metal.\cite{Maier} Even under non-resonant excitation, the spontaneous emission rate can be enhanced by coupling a dipolar emitter in the slot waveguide structure due to strong confinement of electromagnetic modes between the two metal surfaces.\cite{Jun} Large scale spacers of nanometer size gaps can be prepared using Al$_{2}$O$_{3}$ \cite{Chen} and a self-assembled monolayer.\cite{Beesley} The incorporation of quantum dots (QDs) using linker molecule has also been demonstrated.\cite{Tripathi} To reduce the interfacial electron-hole recombination, the thiol (--SH) functional group is used as a hole quencher, which also has the potential to anchor the catalytic center.\cite{Das,Yue,AR} Thiol exhibits a strong affinity towards the transition metal, such as Au, Ag, Pt, and metal chalcogenide-based semiconductors, e.g. CdS and CdSe nanoparticles.\cite{Pham,Love}

Advances in the development of plasmonics have stimulated intensive investigations of electrical detection of plasmon resonance using electrical currents in semiconductors generated due to an energy transfer from plasmons.\cite{Tang,Atwater,Moskowitz} The near-field optical intensity resulting from the interaction of light with nanostructured metals can exceed the incident light intensity by two to three orders of magnitude.\cite{Seo} However, the transformation of this near field optical energy into electricity in semiconductors was given much less attention.\cite{Moskowitz,Falk,Ishi} This allows for easy integration and high-speed, low-capacitance operation of planar devices that converge to optics and electronics.

Here, the plasmon-enhanced photocurrent is observed in thiol-linked-CdSe QDs formed in gold nanogaps. The surface plasmon resonance peaked at about 500 nm is verified by using a photoacoustic (PA) technique. The photoexcited electrons in CdSe transfer to the neighboring dot and then to closely spaced Au surface through thiol links thus forming the photocurrent, which can be tuned by a bias voltage.

\section{Experimental}
\begin{figure}
\includegraphics[width=80mm]{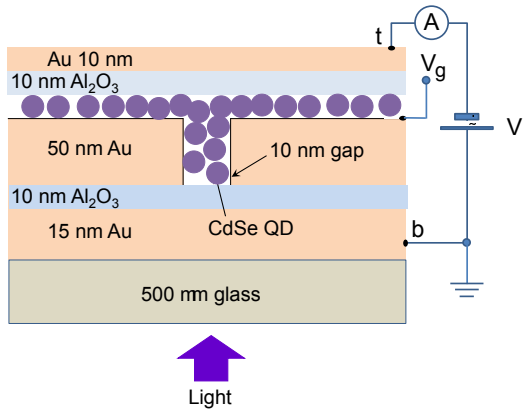}
\caption{\label{Scheme} Schematics of the sample fabrication procedure and photocurrent measurement setup.}
\end{figure}

\begin{figure}
\includegraphics[width=90mm]{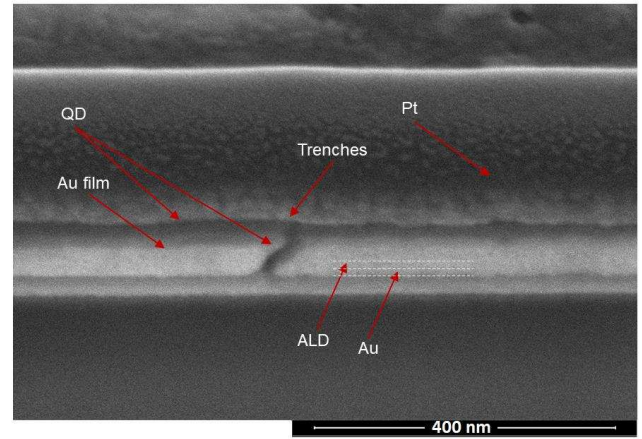}
\caption{\label{SEM} SEM cross-sectional image of the grown structure.}
\end{figure}

A self-assembly lithography method was used to manufacture CdSe QDs-gold nanogap structure with periodic dimensions, as was previously proposed [31,19].\cite{Tripathi2016,Tripathi} The resulting structure is shown schematically in Fig.~\ref{Scheme}. The electrical circuit measuring a current through the structure includes two 10 nm thick layers of Al$_{2}$O$_{3}$. Cross-sectional scanning electron microscopy (SEM) image of the structure is given in Fig.~\ref{SEM}. The breakdown voltage for ALD Al$_{2}$O$_{3}$ is found to be about 0.7 V/nm, so that the 10 nm Al$_{2}$O$_{3}$ layer should work as an insulator for applied voltages V smaller than  5V. The value $V=$5 V was therefore fixed in the photocurrent measurements. A gate bias voltage $V_{g}=\pm$0.1 V was applied to a 50 nm thick Au as shown in Fig.~\ref{Scheme}.

Our measurements demonstrate that a 14 nm thick gold layer has a 50\% transmission of light in the visible region being also suitable to use as a conducting layer. The structure was illuminated from the bottom side, so the light was remarkably blocked by the 50 nm thick gold. Consequently, the photo-induced processes in the CdSe QDs placed in the nanogaps dominate the effects discussed below.

\begin{figure}
\includegraphics[width=70mm]{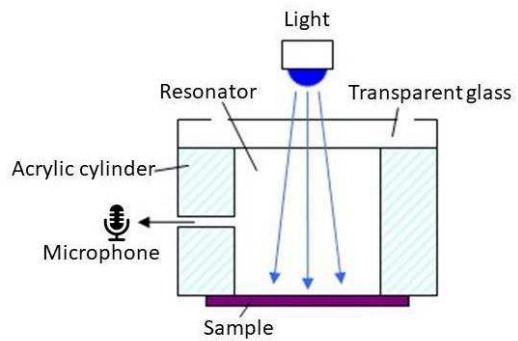}
\caption{\label{PAsetup} Schematic view of the air filled photoacoustic  resonator cell.}
\end{figure}

Schematic view of the photoacoustic resonator cell is shown in Fig.~\ref{PAsetup}. When light absorption occurs in the structure, a resonance-enhanced acoustic signal is generated and sensed by a microphone. The PA measurements were performed in the modulation frequency range from 10 Hz to 1 kHz. A chopped 250 W tungsten halogen lamp (Osram) and a 405-nm pulsed laser diode (LD) were used as light sources. The chopped or pulsed radiation from the external source passed through a grating monochromator and guided to the cell shown in Fig.~\ref{PAsetup}.

\section{Results and discussion}
\begin{figure}
\includegraphics[width=70mm]{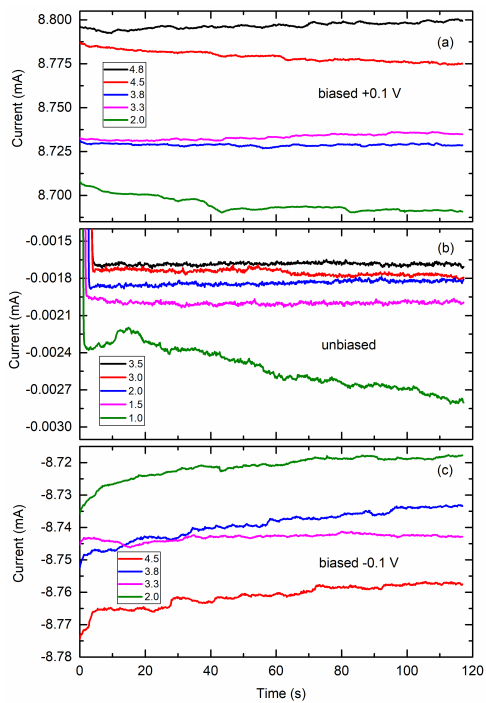}
\caption{\label{PC} Time evolution of the photoinduced current through the structure shown in Fig.~\ref{Scheme} at a broad band excitation with relative intensities indicated in the rectangular areas and $V_{g}=+$0.1 V (a), 0 (b) and --0.1 V (c).}
\end{figure}

Figure~\ref{PC} shows the photocurrent response to the broad band light irradiation, which can be tuned by a bias voltage. It is seen that the photocurrent increases with $V_{g}$ and the current direction changes by reversing the sign of $V_{g}$ (Fig.~\ref{PC}(a)) compared to that observed for the negative potential (Fig.~\ref{PC}(b)) or unbiased middle electrode (Fig.~\ref{PC}(c)).

The Au used for the top and bottom electrodes has a work function of about 5 eV,\cite{Anderson} whereas the electron affinity and bandgap of CdSe QDs are 4.3 eV and 2.2 eV, respectively.\cite{Oertel} The characteristic work functions for Al$_{2}$O$_{3}$ have also been discussed.\cite{Zheng} It can be assumed that the photoexcited electrons in CdSe QDs transfer from the dots to closely spaced middle Au layer through thiol links after they are generated in CdSe. These electrons have energies higher than the Fermi energy of Au, which enables them to transfer to the gold layer directly. Indeed, if an electric field given by $V_{g}$ is applied across the structure, it provides a current flowing through the two Al$_{2}$O$_{3}$ dielectric layers with a thickness of 10 nm, which is usually considered as an upper thickness limit in case of the quantum mechanical tunneling.\cite{Chiu,Molina-Reyes} This forms either the positive or the negative photocurrent directed downwards or upwards in Fig.~\ref{Scheme}, respectively, via electron transport through the Al$_{2}$O$_{3}$ layers, depending on the sign of $V_{g}$. At $V_{g}=$0, the electric field is much smaller and a small negative current is observed in Fig.~\ref{PC}(b), as defined by the polarity of the external voltage applied to the structure.The electric field of incident light induces coherent collective oscillation of free electrons in the metal layer. This is coupled to a positively charged metallic core yielding dipolar oscillations resonant with the incident light at a specific frequency of Au layer. In our structure, the surface plasmon resonance (SPR) wavelength is approximately 500 nm, as confirmed in previous studies.\cite{Tripathi2016,Tripathi} Here, the SPR enhanced photoelectric activity is checked using a photoacoustic  technique, as the resulting acoustic pressure can be related to the absorption coefficient of absorbing medium.\cite{Mandelis}

\begin{figure}
\includegraphics[width=70mm]{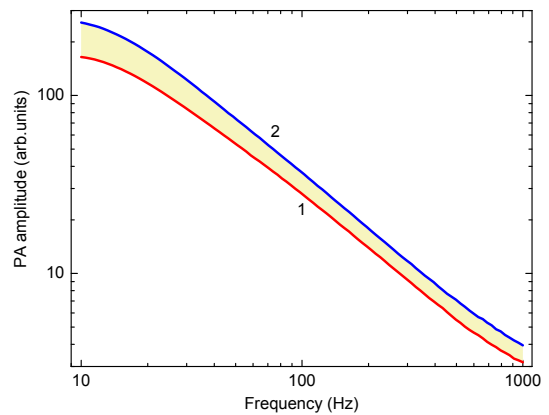}
\caption{\label{PA1} Photoacoustic signal increase due to CdSe quantum dots excited at the wavelength of 405 nm. Curve 1 - before, curve 2 - after filling the nanogaps with CdSe QDs.}
\end{figure}

Figure~\ref{PA1} shows a comparison of the PA signal magnitudes obtained with a 405 nm LD light on the same structure, before (curve 1) and after (curve 2) filling the nanogaps with CdSe QDs. It is seen that the PA magnitude increases by about 20\% due to QDs, whereas the frequency dependence does not change much in the two types of structures.

\begin{figure}
\includegraphics[width=70mm]{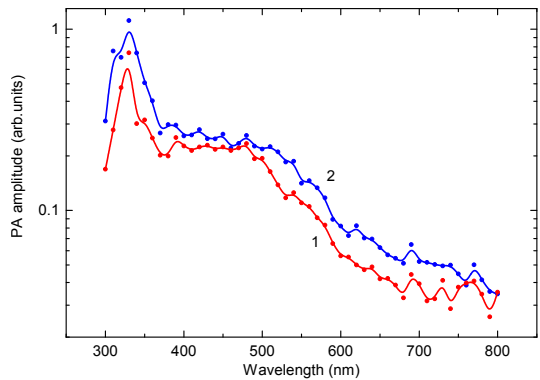}
\caption{\label{PA2} Spectrum of the photoacoustic signal observed before (1) and after (2) filling the nanogaps with CdSe QDs.}
\end{figure}

Figure~\ref{PA2} shows the photoacoustic  spectra of our samples. The enhancement in the PA response peaked at about 500 nm, related to plasmonic absorption by the Au layers (curve 1 in Fig.~\ref{PA2}). The response is found to even further enhance throughout the spectrum due to CdSe QDs (curve 2 in Fig.~\ref{PA2}). The underlying physical mechanisms responsible for the photocurrent behavior observed in Fig.~\ref{PC} may come from the fact that noble metal nanostructures exhibit localized SPR leading to an extremely enhanced localized electromagnetic field near the surface.\cite{Mao} The likely mechanism of the photocurrent in the CdSe QDs-gold nanogap structure is as follows. The quantum dot is excited with above-band gap light to generate electron-hole ($e^{-}-h^{+}$) pairs. While the holes may acquire electrons from impurities adsorbed on the QD surface, moveable excess electrons in CdSe transfer to the neighboring dot and then to closely spaced Au surface through thiol links, establishing a smooth transmission channel of electrons and forming the photocurrent. However, the light absorbance of CdSe QDs is generally weak due to a small number of the dots involved in the absorption processes, so that the photocurrent is small. When the surface of CdSe QDs is in a close proximity with Au, the photocurrent can be enhanced by the localized surface plasmons of gold nanotrenches, e. g. with the bandgap breaking effect.\cite{Zhang}

In this case, the resonance energies of the surface plasmons can be transferred to CdSe QDs non-radiatively and excite the interband transition in CdSe. The power of this energy transfer and the generation rate of the $e^{-}-h^{+}$ pairs are proportional to the square of the local field strength.\cite{Zhang} Consequently, the enhanced localized electric field near the Au surface close to the CdSe QD results in an increased density of the electron-hole pairs. Meanwhile, applying a bias voltage $V_{g}$ makes the photogenerated $e^{-}-h^{+}$ pairs separate rapidly, which precludes their recombination and hence increases the photocurrent, as indeed observed in Fig.~\ref{PC}(a) and (c) in comparison with the unbiased case shown in Fig.~\ref{PC}(b).

\section{Conclusions}

In summary, low voltage-controlled, plasmon-enhanced photocurrent is observed in a system of thiol-linked CdSe quantum dots grown in Au nanogaps. The surface plasmon resonance peak at about 500 nm is revealed using a photoacoustic  technique. The observed enhancement in the photoacoustic  response is related to plasmonic absorption by the Au layers and the response is further enhanced by about 20\% due to CdSe QDs. The photocurrent behavior is explained by enhanced localized electric fields near the Au surface. The photoexcited electrons in CdSe transfer to the neighboring dot and then to closely spaced Au surface through thiol links. The plasmonic nanogap structure geometries can open up new possibilities for carrier collection from quantum-dot photovoltaic devices, in nanometer-scale photodetectors and sensors.

\section{Acknowledgements}

This work was supported by the National Research Foundation of Korea (NRF) grant funded by the Korean government (MSIP: NRF-2015R1A3A2031768, NRF-2017R1E1A1A01074650) and the research project funded by start-up (1.190055.01) and U-K Brand (1.190109.01) of UNIST (Ulsan National Institute of Science and Technology). The work at Kyiv was funded by the Ministry of Education and Science of Ukraine, grant numbers 0119U100303 and 0122U001953. O.K. acknowledges support from the Erwin Schr\"{o}dinger Institute by the Special Research Fellow Programme.

\end{document}